\documentstyle[12pt]{article}
\makeatletter
\newdimen\normalarrayskip              
\newdimen\minarrayskip                 
\normalarrayskip\baselineskip
\minarrayskip\jot
\newif\ifold             \oldtrue            
\def\arraymode{\ifold\relax\else\displaystyle\fi} 
\def\eqnumphantom{\phantom{(\theequation)}}     
\def\@arrayskip{\ifold\baselineskip\z@\lineskip\z@
     \else
     \baselineskip\minarrayskip\lineskip2\minarrayskip\fi}
\def\@arrayclassz{\ifcase \@lastchclass \@acolampacol \or
\@ampacol \or \or \or \@addamp \or
   \@acolampacol \or \@firstampfalse \@acol \fi
\edef\@preamble{\@preamble
  \ifcase \@chnum
     \hfil$\relax\arraymode\@sharp$\hfil
     \or $\relax\arraymode\@sharp$\hfil
     \or \hfil$\relax\arraymode\@sharp$\fi}}
\def\@array[#1]#2{\setbox\@arstrutbox=\hbox{\vrule
     height\arraystretch \ht\strutbox
     depth\arraystretch \dp\strutbox
     width\z@}\@mkpream{#2}\edef\@preamble{\halign \noexpand\@halignto
\bgroup \tabskip\z@ \@arstrut \@preamble \tabskip\z@ \cr}%
\let\@startpbox\@@startpbox \let\@endpbox\@@endpbox
  \if #1t\vtop \else \if#1b\vbox \else \vcenter \fi\fi
  \bgroup \let\par\relax
  \let\@sharp##\let\protect\relax
  \@arrayskip\@preamble}
\def\eqnarray{\stepcounter{equation}%
              \let\@currentlabel=\theequation
              \global\@eqnswtrue
              \global\@eqcnt\z@
              \tabskip\@centering
              \let\\=\@eqncr
              $$%
 \halign to \displaywidth\bgroup
    \eqnumphantom\@eqnsel\hskip\@centering
    $\displaystyle \tabskip\z@ {##}$%
    &\global\@eqcnt\@ne \hskip 2\arraycolsep
         $\displaystyle\arraymode{##}$\hfil
    &\global\@eqcnt\tw@ \hskip 2\arraycolsep
         $\displaystyle\tabskip\z@{##}$\hfil
         \tabskip\@centering
    &{##}\tabskip\z@\cr}
\makeatother

\def\beq{\begin{equation}}
\def\eeq{\end{equation}}
\def\bea{\begin{eqnarray}}
\def\eea{\end{eqnarray}}

\def\stackreb#1#2{\mathrel{\mathop{#2}\limits_{#1}}}
\def\theequation{\arabic{equation}}  

\begin{document}

\begin{titlepage}
\setcounter{footnote}0
\begin{center}
{\it P.N.Lebedev Institute preprint} \hfill FIAN/TD-8/92\\
{\it I.E.Tamm Theory Department} \hfill hepth@xxx/92\#9208022
\begin{flushright}{June 1992}\end{flushright}
\vspace{0.1in}{\Large\bf On string field theory for $C \leq 1$}
\footnote{\it to appear in Proceedings of 16th Jonhs Hopkins Workshop
on Current Problems in Particle Theory: Pathways to Fundamental Theories}
\\[.4in]
{\large  A.Marshakov}\\
\bigskip {\it Theory Department \\  P.N.Lebedev Physics
Institute \\ Leninsky prospect, 53, Moscow, 117 924, Russia},
\footnote{E-mail address: tdparticle@glas.apc.org}\\ \smallskip
\end{center}
\bigskip

\centerline{\bf ABSTRACT}
\begin{quotation}
\noindent
I present a short review of our results with S.Kharchev, A.Mironov, A.Morozov
and A.Zabrodin on Generalized Kontsevich model which in a sense can be
interpreted as unifying ``string field theory" for  $c < 1$  minimal series
coupled to 2d gravity. The problem of interpolation between different models is
discussed. It is found that this problem is closely connected with
``deformations" within the set of solutions to KP hierarchy, described by a
sort of reparameterization of a spectral curve and change of asymptotics of the
basis in the Grassmannian. The  $c \rightarrow  1$  limit is
considered along this line.
\end{quotation}
\bigskip\bigskip
\end{titlepage}
\newpage

1.   Recently [1,2,3] a new model was proposed which describes a
non-perturbative solution to the  $c < 1$ $2d$ conformal matter coupled to
the two-dimensional gravity. It seems now that this model could be considered
as
the first successful attempt of constructing a {\it string field theory}
or effective theory of string models. It is necessary to point out from the
beginning that by string field theory we would mean more than a conventional
definition as a field theory of functionals defined on string loops - it must
rather mean a sort of effective theory which gives all the solutions to
classical string equations of motion (2d conformal field theories coupled to 2d
gravity) as its vacua and allows us to consider all of them on
equal footing (within the same Lagrangian framework) and maybe even describe
the
flows betveen different string vacua. Of course, it should reproduce
perturbation expansion around any of these vacua.
In this sense, the conventional string field theory was not true
effective model, because it contained a {\it fixed} set of variables which
correspond to a concrete vacuum (say, 26 free scalar fields). Therefore, it
doesn't have even {\it a priori} a possibility to make a flow to another
classical solution (maybe only except for some simple change of a background),
$i.e$. conventional string field theory might describe only some small
perturbation around given classical solution in terms of the coordinates
equivalent to the matter variables in the Polyakov path integral.

The other side of the problem is
connected with the non-perturbative definition of string theory. In
conventional
approach (critical string) even the perturbative expansion was ill-defined due
to the presence of tachyon in the spectrum or in other words due to the
instability of the classical solution.  The only chance to get a sensible
effective theory appears after we make sense to the non-perturbative
description. Such description appeared (and exists now only in the case of
non-critical and moreover ``non-tachyonic" strings) in the language of ordinary
matrix models after the double scaling limit has been found [4-7]. The deep
connection between matrix models and integrable theories [8] made possible
``axiomatic" description of string field theory which was actually formulated
in
[9] by three following statements:

(i)
\beq
{\cal F}(T) = \log \ \tau (T)
\eeq
$i.e$. the partition function (or better generating function) for string
correlators is given by logarithm of {\it tau}-{\it function} of the
Kadomtsev-Petviashvili (KP) hierarchy

(ii)
\beq
{{\cal L}_{-1}\tau (T)\over \tau (T)} = 0
\eeq
or the particular choice of the KP-solution is determined by so-called
{\it string equation}.

The notation  ${\cal L}_{-1}$ means that the string equation is the first from
the tower of (extended) Virasoro constraints which vanish the  $\tau $-function
reproducing the partition function (generating function for all the
correlators) in the non-perturbative string theory acting as differential
operators with respect to the infinite set of KP times $\{T\} = T_1,T_2,...$

(iii) Depending on particular {\it reduction} of the KP hierarchy and
the particular form of string equation $and/or$ (extended) Virasoro algebra one
can describe the series of conformal matter plus 2d gravity where the matter
central charges are given by
\beq
c_{matter} = c_{p,q} = 1 - 6{(p-q)^2\over pq}
\eeq
with one of the coprime numbers, say  $q$  is fixed and determines the
particular $q$-reduction of the KP hierarchy (e.g. $q=2$  corresponds to the
Korteweg de Vries (KdV) hierarchy,  $q=3$  - to the Boussinesq hierarchy, and
in general fixed finite  $q$-reduction  is called $q$-th KdV) and  $p$  is
arbitrary.

Of course, the central charges (3) are known as the central charges of minimal
conformal models [10] which can be described (before coupling to gravity)
as highly reduced  $c = 1$ $2d$
free-field theories [11,12]. That means that what we know up to now is the only
case of ``highly-noncritical" string models where the total matter-gravity
central charge
\beq
c_{matter} + c_{gravity} = 26
\eeq
is ``dominated" by the contribution of 2d gravity. This is far from the case of
critical string $(c_{matter} = 26)$  and close to the case of pure
two-dimensional gravity. Unfortunately, up two now this is the only region
where it is possible to formulate string theory consistently and at least put
the question what is the internal principle which might allow one to choose
dynamically a string vacuum.

The idea to play with highly reduced theories leads automatically to a
conclusion that the effective string field theory should not necessarily be a
{\it field theory} in a literal sense, and this is at least naively consistent
with the result, being presented in a form of (not really functional)
matrix integral. Moreover, one can actually consider this matrix integral as a
{\it finite} one, because the result in fact is independent of the size  $N$
of matrices. Indeed, the only place where this parameter appears is the
definition of coupling constants in terms of  $N$  Miwa ``spectral parameters";
see below). The fact of  $N$-independence in the theory could be interpreted as
a {\it cutoff independence}  of topological models, though the field
theoretical sense of Miwa transform is still not clear.

Another crucial role of the Miwa coordinates is connected with the problem of
interpolation between different string vacua. We know, that in terms of the KP
times themselves there is no good flow between various critical points of usual
matrix models (see for example [13]). In a sense this means that in terms of KP
time-variables certain limits of them to zero values could be singular
\footnote{These ``non-vanishing" parameters were also found in continuum
formulation (see, for example, N.Seiberg's contribution to this volume.}
. Introducing Miwa's variables allows us in principle to go around this
problem,
considering flows as certain reparametrizations of spectral curve, which in the
language of hierarchies of integrable equations is connected with so-called
equivalent hierarchies [14,15].

\bigskip
\noindent
2.   The Lagrangian description of the effective  $c < 1$  string theory is
based on the equivalence of the topological and quantum 2d gravity [16,17,1,2].
In other words the solution to the topological 2d gravity found by Kontsevich
[18] is equivalent to the particular solution of the KdV hierarchy obeying
Virasoro constraints or string equation (${\cal L}_{-1}$ -- constraint) (2)
with
\beq
{\cal L}_n = {\cal L}^{(2)}_n = {1\over 2}\sum _{k \ \ odd} kT_k
{\partial \over \partial T_{k+2n}} +
{1\over 4}\sum _{a+b=2n}{\partial ^2\over \partial T_a\partial T_b} +
$$
$$
+ {1\over 4}\sum _{a+b=-2n}aT_abT_b + {1\over 16} \delta _{n,0} -
{\partial \over \partial T_{3+2n}}
\eeq
being second order differential Virasoro operators acting to the KdV-hierarchy
$\tau $-function, depending only (in this particular case of  $q=2$  reduction)
upon odd times
\beq
\tau _{KdV}(T) = \tau _{KdV}(T_1,T_3,...)
\eeq

This is the case of ``pure gravity", $i.e$.  $c$  is given by (3) with  $q=2$,
$p=2k+1  (k \geq  0)$. The idea is that the solution to the eqs.(5), (6) can be
represented in the form of matrix integral
\medskip
\beq
{\cal Z} [M] = {{\int DX\ \exp  -Tr(MX^2+X^3/3)} \over
{\int DX \ \exp -TrMX^2}} =
$$
$$
= C[M]\int   DX\ \exp \ Tr(M^2X - X^3/3)
\eeq
where  $X \equiv  \|X_{ij}\|$  and  $M = \|M_{ij}\| - N\times N$ hermitean
matrices (the last one can be taken to be diagonal because ${\cal Z} [M]$
depends only on its eigenvalues) and the normalization  $C[M]$  is given by
\beq
C[M] = \exp (- {2\over 3}TrM^3) \det (M^T\otimes I+I\otimes M)^{1/2}
\eeq

The partition function  $Z[M]$  can be decomposed in series over
\beq
T_n = {1\over n}TrM^{-n} = {1\over n}\sum ^N_{j=1}\mu ^{-n}_j
\eeq

($n$ odd) and the coefficients reproduce the intersection indices on module
spaces of Riemann surfaces with punctures [18] or the (1,2) (topological)
gravity theory with  $c_{matter} = c_{1,2} = -2$. In terms of times (9) the
partition function (7) gives a representation for $\tau $-function of KdV
hierarchy obeying Virasoro constraints (5). This can be proven directly using
the properties of matrix integral (7) [17].
Kontsevich's potential  $V(X) = X^3/3$  is the simplest one and
describes the $q=2$ series of ``pure gravity". However, the generalization to
higher series is straightforward [1,2], and is given by (7) with an
arbitrary polynomial potentials (non-polynomial potentials would rather
correspond to less trivial cases both from ``stringy" and ``integrable" point
of
view, see below)
\beq
{\cal Z}_{GKM}[M,V] = {{\int DX\ \exp  -Tr\ U(M,X)} \over
{\int DX\ \exp  -TrU_2(M,X)}} =
$$
$$
= C[M|V]\int   DX\ \exp \ Tr[V'(M)X - V(X)]
\eeq
with
\beq
C[V|M] = \exp \ Tr[V(M) - MV'(M)] \det [V''(M)]^{1/2} {\Delta (V'(M)) \over
\Delta (M)}
$$
$$
U(M,X) = V(M+X) - V(M) - XV'(M)
$$
$$
U_2(M,X) =
\lim_{\epsilon \rightarrow 0}
{1\over \epsilon ^2} U(M,\epsilon X)
\eeq
which is nothing but a sort of ``effective potential" for matrix theory. For
the case  $V(X) = X^3/3$ \ \  eqs. (10), (11) reduce to (7), (8). Two general
statements (i), (ii) are easily derived from the matrix-integral
representation, first is that
\beq
{\cal Z}[T(M)|V] \equiv  \exp \ {\cal F}[T(M)|V] =
$$
$$
= {\det _{(ij)}\Phi ^{\{V\}}_i(\mu _j)\over \Delta (\mu )} = \tau [T_n =
{1\over n}TrM^{-n}|V]
\eeq
with
\beq
\Phi ^{\{V\}}_i(\mu ) = \exp  [V(\mu ) - \mu V'(\mu )] V''(\mu )^{1/2}\times
$$
$$
\times \int   dx\ x^{i-1}\exp \ Tr[V'(\mu )x - V(x)]
\eeq
where the determinant formula (12) means that the partition function satisfies
the Hirota difference bilinear relation in Miwa coordinates (9) [19] and thus
is a $\tau$-function of KP hierarchy while the second reads
\beq
{{\cal L}^{\{V\}}_{-1}{\cal Z}[T(M)|V]\over {\cal Z}[T(M)|V]} = 0
\eeq
where
\beq
{\cal L}^{\{V\}}_{-1} =
\sum _{n\geq 1}Tr[V''(M)M^{n+1}]^{-1}\partial /\partial T_n +
$$
$$
+ {1\over 2}\sum _{i,j}{1\over V''(\mu _i)V''(\mu _j)} {V''(\mu _i) -
V''(\mu _j)\over \mu _i - \mu _j}  - \partial /\partial T_1
\eeq
(see [2] for details). Equations (14), (15) mean that for {\it any} (at least
polynomial) potential  $V(X)$  we get from GKM a $\tau $-function of KP
hierarchy which satisfies string equation, so, at least naively we can preserve
both these properties (integrability and string equation) varying the potential
$V(X)$  smoothly between, say, two monomials corresponding to particular
$(p=fixed,q)$ series  (monomial  $X^{p+1}/(p+1)$  gives $(p,q)$ solution with
{\it fixed $p$} in terms of $p$-th KdV reduction of KP hierarchy). In this
sense we immediately obtain a string field theory (in the sense of sect.1)
description of all discrete series with  $c<1$  coupled to two-dimensional
gravity.

However this is not true exactly due to two important things which are now in
order. First one is connected with the choice of a particular ``critical point"
within one series with {\it fixed $p$}, and the second one concerns flows
between two different  $p$'s or two different classes of potentials
\footnote{Of course, these two are actually just the same problem due to $p-q$
duality, which is however not manifest in the language of matrix models (see,
for example [20]).}.

Naively Generalized Kontsevich model gives us a ``topological" solutions within
$c_{p,q}$ series - $i.e$. only points where  $(p,q) = (p,1)$  $c = c_{p,1} = 1
-
6{(p-1)^2\over p}$ \ \
\footnote{Again, even at the level of 2d conformal field theory these are
singled points (with negative integer central charges, integer dimensions
primary fields etc); see also P.West's contribution to this volume.}
. This is determined by a simple fact that $\tau $-functions are defined as
formal series in times $\{T\}$ or in other words for  small $T$'s and this
corresponds to the limit  $M \rightarrow  \infty $  in terms of spectral
parameter  $(\mu _j \rightarrow  \infty $  altogether). The particular critical
point is determined by the following constraints [9]
\beq
\hat T_2 = \hat T_3 = ..\hbox{. } = \hat T_{p+q-1} = 0
$$
$$
\hat T_{p+q} = const \neq  0
\eeq
where for GKM we defined $(V(X) \equiv  \sum v_kX^k)$
\beq
\hat T_n = \hat T^{(V)}_n = T_n - (n-1)v_n
\eeq
(17)
(in particular for monomial potential  $v_n = {\delta _{n,p+1}\over n}$,
$\hat T_n = T_n - {n-1\over n} \delta _{n,p+1})$  and the shift (17) is
determined by the requirement of absorption of linear (like
$\partial /\partial T_1)$ terms in the expressions for Virasoro generators,
$i.e$. by the exact coincidence with the constraints for ordinary matrix models
[9,21]. Thus, we see that for any monomial potential in GKM one gets
\beq
\hat T_n = - {n-1\over n} \delta _{n,p+1} + O(1/M^n)
\eeq
$i.e$.  $\hat T_{p+q} \neq  0$  only for  $q=1$, and the first correction in
$1/M$  will be given by  $T_1$. For ``pure" Kontsevich's case  $V(X) = X^3/3$
we obtain in such a way the theory with  $c = c_{2,1} = -2$  [22,23] which
doesn't correspond to any critical behaviour of the Hermitean 1-matrix model
which starts from  $q=3$, $\hat T_5 \neq  0$, $c_{2,3} = 0$ (pure gravity),
$q=5$, $\hat T_7 \neq  0$, $c_{2,5} = - 22/5$ (Yang-Lee model) {\it etc}.

In order to get higher critical points one should try to consider more
complicated choices for matrix  $M$, or in other words to consider expansion
near different from infinity points on the surface of spectral parameter. For
example, instead of expansion over  $Z = 1/M$  near  $Z = 0$  we can take
for the simplest case  $V(X) = X^3/3$  the following choice
\beq
Z = {1\over 3}\left[
\begin{array}{ccc}
{z\omega _1+\epsilon }& & \\
 & {z\omega _2+\epsilon} & \\
 & & {z\omega _3+\epsilon}
\end{array}
\right]
\eeq
$i.e$. take matrix  $Z$  in block form where any block  $z$  is multiplied by
corresponding root of unity  $\omega _k = \exp ({2\pi ik\over 3})$. Then
obviously
\beq
\hat T_1 = Tr\epsilon
$$
$$
\hat T_3 = Trz^3 - {2\over 3} + Tr\epsilon ^3
$$
$$
\hat T_5 = 10Trz^3\epsilon ^2 + ...
$$
$$
\hat T_7 = 7Trz^6\epsilon  + ...
$$
$$
\hat T_9 = Trz^9 + O(\epsilon ^3)
\eeq
and if we adjust  $Trz^3 = {2\over 3}$,  then the first non-zero time will be
$\hat T_9$ (we remind that the case of cubic potential corresponds to the KdV
reduction of KP hierarchy and the partition function is independent of all even
times  $T_{2n}$, in particular of  $T_6)$. There will be other non-zero times
(say, $\hat T_{27})$ but they can be switched off by tunning proper behavior in
the limit  $N \to \infty$ and the important one is  $\hat T_9$, moreover the
critical behaviour is determined by the lower degree term on sphere
\beq
\partial /\partial T_1{{\cal L}^{(2)}_{-1}\tau _{KdV}\over \tau _{KdV}}
=\sum _{k\geq 1}(2k+1)\hat T_{2k+1}u^k- {1\over 8}\hat T_1+ O(\partial u,...)=0
\eeq
so that
\beq
u \sim  T^{1/k}_1
\eeq
where  $\hat T_{2k+1}$ is the {\it first} non-vanishing term. The behaviour is
determined by  $2k+1 = p+q$, or

\beq
{1\over k} = - {2\over p+q-1} = \gamma _{str}
\eeq
The other problem is that now  $\hat T_7$ is of the same order as  $\hat T_1$,
but this should be changed by usual renormalization (dependent upon a
particular critical point) in string equation.

Of course, the above example only touches the whole problem and is not
satisfactory, because, for example, it doesn't lead to the most interesting
point of pure gravity. However, it demonstrates the main idea --
already this flow in ``$p$-direction" is connected with a specific
reparametrization of spectral curve and change the asymptotics of basis vector
in the Grassmannian  (now  $z_j = 1/\mu _j$ have different limits for different
$j$, and expansion should be taken in different points). Below, we'll see that
that actually the same phenomenon appears when we consider other problems of
formulating string field theory.

\bigskip
\noindent
3.   Now let us pass to interpolation between various series with different
$p$'s (flows in ``$p$-direction"). The example can be given by the potential
\beq
V(X) = \alpha  {X^{k+1}\over k+1} + \beta  {X^k\over k}
\eeq
with  $\alpha  + \beta  = 1$. This should describe the flow between points with
$\alpha  = 0$  and  $\alpha  = 1$  which is nonsingular everywhere except for
the place of choosing of integration contour. In principle any matrix integral
of the type (10) is determined by analytical continuation (or contour
deformation) from a conventional definition of corresponding (generalized) Airy
function. For various monomials in the exponent the definitions of such
contours are different and the addition even of a small piece like that in (24)
can change the contours drastically. However, this is nothing but the same sort
difficulty as in the ordinary non-perturbative definition of field theory path
integral (with the same potential (24)) and it is not too serious problem. The
deformation (24) around monomial potential by lower order terms is described
by so called
Landau-Ginsburg flows and it uses the fact that the derivatives with respect to
the first  $k \leq  p$  times are related to the insertion of corresponding
monomials (see [24] for details)
\beq
{\partial {\cal Z}\over \partial T_k} = \langle TrM^k -
TrX^k \rangle , \ \
k = 1,...,p
\eeq
(${\cal Z} \equiv  \langle 1 \rangle $). Moreover, the flow around a given
``critical point" (determined by the higher-degree term in potential;  $k+1$ -
in (24)) can be at least partially absorbed into the definition of times via
the change of spectral parameter
\beq
\tilde M = [V'(M)]^{1/k}
\eeq
(a sort of ``positive" Virasoro reparameterization -- not moving the point of
expansion, in contrast to the case considered in sect.2 above) and the
redefined
partition function
\beq
\tilde {\cal Z}[\tilde T|\tilde V] = {C[\tilde M|\tilde V]\over C[M|V]}
{\cal Z}[T|V]
\eeq
is still a $\tau $-function of $k-reduced$ KP hierarchy in terms of new
$\tilde T_n = {1\over n}Tr\tilde M^{-n}$ variables.

However, the limit  $\alpha  \rightarrow  0$  will be a singular one. It is
exactly the case when the contour jumps, or in other words one has to change
the region of definition of spectral parameter. The other thing is that the
string equation is deformed smoothly in terms of Miwa coordinates but has two
absolutely different expansions in times  $T_n$. Another immediate consequence
of (26) is different role of times  $T_n$ for  $n\leq p$  and  $n>p$. In the
language of topological theories the first ones correspond to so called primary
fields while the second to their descendants (see, for example, [25] and
references therein).

The ``Landau-Ginsburg" deformation (27) doesn't even change the ``critical"
point of concrete solution, because in new  $\tilde T$-variables the solution
has the same reduction. It corresponds rather to an infinitesimal deformation
of a GKM in the vicinity of a critical point. The corresponding change of
spectral parameter or spectral curve reparameterization (26) is also
infinitesimal in the sense that it doesn't move (in contrast to the previous
section) the point of expansion (and even the asymptotics of basis vectors). In
the language of integrable hierarchies this corresponds to so called
{\it equivalent hierarchies} [14,15], where new and old times are connected by
a triangular linear transformation, and the potentials of one solution are
functionals of the potentials of another one. The main feature of equivalent
hierarchies is that not any transformation of times but only those, induced by
spectral reparametrizations like (26). We demonstrated above that in more
general situation we have a more complicated case of spectral reparametrization
(as well as when considering  $c \rightarrow  1$  limit below) but this is not
yet formulated in terms of the properties of the hierarchy of integrable
equation. We are going to return to this problem elsewhere [26].

\bigskip
\noindent
4.   Now, let us discuss the $c \rightarrow  1$  limit of Generalized
Kontsevich
model. The exact way to do this is to take  $p \rightarrow  \infty $  limit
keeping  $q = p + 1$  (or at least difference  $p - q$  fixed and finite, then
$c_{p,p+1} = 1 - 6/p(p+1) \stackreb{p \rightarrow \infty }{\rightarrow} 1)$.
Unfortunately, this is hard to perform exactly (due
to all mentioned above problems) let us instead try to analyze the naive limit
$p \rightarrow  \infty $  in the model with monomial potential  $V(X) =
X^{p+1}/(p+1)$. Changing the variables
\beq
Y = - {1\over p+1}\Lambda ^{p+1}X^{p+1}
\eeq
in the integral
\beq
\int   DX\ \exp \ Tr[\Lambda X - X^{p+1}/(p+1)]
\eeq
one gets
\beq
\int \hbox{  DY }\ \exp  \{Tr[- (p+1)Y^{1/(p+1)} + \tilde \Lambda Y] +
\log {\partial (X)\over \partial (Y)}\}
$$
$$
\tilde \Lambda  = (-p-1)^p \Lambda ^{-(p+1)}
$$
$$
T_n = {1\over n} Tr\Lambda ^{-n/p} \sim  {1\over n}
Tr\tilde \Lambda ^{n/p(p+1)}
\eeq
where the last term stands for Jacobian of the transformation (28). This
Jacobian can be determined from
\beq
DY \sim  (det\Lambda ^{p+1})^N D(X^{p+1})
\eeq
and
\beq
D(X^{p+1}) = (p+1)^N\hbox{ (detX})^p\left( \prod _{i<j} \sum _{a+b=p}
x^a_i x^b_j\right) ^2DX
\eeq
where  $N$ -- the size of the matrix $X$, and  $\{x_i\}$ - its eigenvalues.
(Eq.
(32) easily follows from decomposition  $X = \Omega^{\dag} x \Omega$,
${X^{p+1}
= \Omega^{\dag} x^{p+1} \Omega}$,  where  $x = diag(x_1,...,x_N)$), and
\beq
DX = D\Omega^{\dag} D\Omega \prod_i dx_i \Delta ^2(x)
\eeq
with  $\Delta (x) = \prod _{i<j}(x_i - x_j)$ -- Vandermonde determinant). It
means that
\beq
DX \sim  DY \left( (\hbox{detX})^p\prod _{i\neq j}\sum _{a+b=p}
x^a_i x^b_j\right) ^{-1} \sim  DY\ \exp  [- {p\over p+1}\hbox{TrlogY -}
$$
$$
- \sum _{i\neq j}\log \sum_{a+b=p}\lambda ^{-a}_iy^{a/(p+1)}_i
\lambda ^{-b}_jy^{b/(p+1)}_j] \sim
$$
$$
\sim  DY\ \exp  [- {p\over p+1}\hbox{TrlogY } -
\sum _{i\neq j}\log \sum_{a+b=p}(\tilde \lambda _iy_i)^{a/(p+1)}(
\tilde \lambda _jy_j)^{b/(p+1)}]
\eeq
Finally, in the limit  $q \rightarrow  \infty $ the integral (30) turns to be
\beq
\int \hbox{  DY }\ \exp  [Tr\tilde \Lambda Y - 2TrlogY - \sum _{i\neq j}\log
{\tilde \lambda _iy_i - \tilde \lambda _jy_j\over \log (\tilde \lambda _iy_i
/\tilde \lambda _jy_j)}] \sim
$$
$$
\sim  \int   D\Xi \ \exp  [Tr\Xi  - 2Trlog\Xi  - \sum _{i\neq j}\log  {\xi _i -
\xi _j\over \log (\xi _i/\xi _j)}]
\eeq

So, we see that in such limit  $p \rightarrow  \infty $  one gets a
construction similar to Penner model [27,28] and all the nontrivial ``Miwa"
times decouple from (35) (this is consistent with (30) which stands that all
nontrivial times are tend to zero in such limit). Thus, in naive limit we can
get nothing in addition to effective theory for puncture operator (tachyon with
zero momenta) which decouples from rest of the theory in such limit (see for
example [29] and references therein).

However, the determinant form of Penner model partition function implies
already
that for fixed values of times it is a Toda lattice tau-function in the sense
of [3] and allows us to
apply to this case the main idea of [3] -- the Toda theory
representation for Generalized Kontsevich models. Indeed, the solution to
the Penner model
\beq
{\cal Z} \sim  \det \ {\cal H}^{(\alpha )}_{ij}
\eeq
with
\beq
{\cal H}^{(\alpha )}_{ij} = \Gamma (\alpha +i+j-1)
\eeq
is nothing but a specific case of GKM.

The solution to generic Toda lattice hierarchy looks like
\beq
\tau _n[T] = \det _{(ij)}H_{i+n,j+n}[T]
\eeq
where matrix elements  $H_{ij}$ satisfy the following equations
\beq
\partial H_{ij}/\partial T_p = H_{i,j-p}\hbox{ for ``positive times" }  T_p
$$
$$
\partial H_{ij}/\partial T_{-p} = H_{i-p,j}\hbox{ for ``negative times" }
T_{-p}
\eeq
($n$  being integer-valued ``zero-time"). In particular, for generalized
Kontsevich model the solution of (38), (39) has the form
\beq
H_{ij}[T_{-p},T_p] = \sum _{k\leq i,l\geq -j}P_{i-k}[T_{-p}] h_{kl}
P_{l+j}[T_p]
\eeq
with
\beq
h_{kl} =
\oint \Phi ^{\{V\}}_k(z)z^l
\eeq
where  $\Phi ^{\{V\}}_k(z)$  are defined in (11)
\footnote{The most interesting example of non-trivial Toda generalization of
GKM is given by a discrete Hermitean matrix model [3,30] which corresponds to
$Tr(\Lambda X - X^2/2 + nlogX)$  in the exponent, ($n$  being Toda zero-time),
when for $n=0$  the GKM integral is Gaussian and thus trivial.}.

Now one can easily introduce positive- and negative-times dependence in (37)
according to (40) and then reconstruct  $\Phi ^{\{V\}}_k(z)$  from (41)
\footnote{C.Vafa told me that negative Toda times in order
to describe  $c = 1$  theory by means of GKM were also used by R.Dijkgraaf
and G.Moore.}
. Indeed,
\beq
h^{(\alpha )}_{ij} = {\cal H}^{(\alpha )}_{ij} = \Gamma (\alpha -1+i+j) =
$$
$$
= \int ^\infty _0{dy\over y} e^{-y} y^{\alpha -1+i+j} =
\oint \phi ^{(\alpha )}_i(z)z^j
\eeq
immediately gives
\beq
\phi ^{(\alpha )}_i(z) = \int ^\infty _0{dy\over y} e^{zy-y} y^{\alpha -1+i}
\eeq
which is a sort of GKM-like representation. The difference with more common
situation for  $c<1$  is in the definition of the contour in (43) and also in
the fact that  $z$-dependence is trivial in (43), because integral is easily
taken with the result
\beq
\phi ^{(\alpha )}_i(z) = {\Gamma (\alpha +i)\over (z-1)^{\alpha +i}} \equiv
\phi _{\alpha +i}(z)
\eeq
and
\beq
\left( {\partial \over \partial z}\right) ^j\phi ^{(\alpha )}_i(z) =
(-)^j\phi ^{(\alpha )}_{i+j}(z) = (-)^j
{\Gamma (\alpha +i+j)\over (z-1)^{\alpha +i+j}}
\eeq
Introducing negative times according to [3], one gets
\medskip
\beq
\phi ^{(\alpha )}_i(z|T_{-p}) =
z^{-\alpha }\exp \left( -\sum _{p>0}T_{-p}z^{-p}\right) \sum  _k
P_k[T_{-p}]\phi ^{(\alpha )}_{i-k}(z)
\eeq
where  $P_k[T_p]$   are Schur polynomials  $(\exp \ \sum T_pz^p =
\sum \ z^nP_n[T_p]).$ or simply
\beq
Z_{c=1} \sim \int DY \exp {Tr ZY + \alpha Tr log Y +
\sum_{k>0} T_{-k} Tr Y^{-k}}
\eeq
with
\beq
T_{+k} = {1 \over k} Tr Z^k
\eeq

Now, we see again that the formulas (44) looks like basis vector of a trivial
element of Grassmannian (sphere) but in a shifted point of spectral curve
(expansion near  $z = 1$). Thus, we conclude that the same phenomena which
corresponds to flows between various pq-solutions is also valid when taking  $c
\rightarrow  1$  limit of GKM. This seems to be a generic feature of $c \leq
1$  effective string theory.

\bigskip
\noindent
I tried to demonstrate above that matrix models in general and especially the
Generalized Kontsevich model give at the moment the most advanced understanding
of  $c \leq  1$  non-perturbative string theory. This is mostly connected with
the appearance of a non-trivial dynamics over coupling constants in this
formulation, described by integrable equations of KP (or Toda lattice)
hierarchy. The particular stringy solutions to KP can be described in a form of
matrix integrals, having the sense of effective string field theory. The
central point is introducing specific coupling by a matrix of Miwa spectral
parameters, thus making the conventional flows in the space of coupling
constants to be a sort of ``spectral flows". From the point of view of
integrable models themselves these spectral flows correspond to equivalent
hierarchies, which have not yet been investigated a lot. We are going to return
to all these problems elsewhere.

\bigskip
\noindent

I am grateful to I.Kogan, C.Vafa, A.Zabrodin and especially to S.Kharchev,
A.Mironov and A.Morozov for very useful discussions. I am grateful for
hospitality to the Lyman Laboratory of Harvard University, Physics Departments
of University of British Columbia and California Institute of Technology where
different parts of this work have been done. I would like also to thank
L.Brink, A.Hofling, R.Marnelius and other organizers of the 16th Johns Hopkins
workshop for very nice and interesting time while in Gothenburg.
\bigskip

\newpage
\centerline{REFERENCES}
\noindent
1. S.Kharchev et.al.  Phys.Lett. {\bf B275} (1992) 311.

\noindent
2. S.Kharchev et.al.  Nucl.Phys.{\bf B380} (1992) 181.

\noindent
3. S.Kharchev et.al.  {\it Generalized Kontsevich model versus Toda hierarchy
and discrete matrix models}  Preprint $FIAN/TD-03/92 (hepth@xxx/9203043).$

\noindent
4. V.Kazakov  Mod.Phys.Lett. {\bf A4} (1989) 2125.

\noindent
5. E.Brezin, V.Kazakov  Phys.Lett. {\bf B236} (1990) 144.

\noindent
6. M.Douglas, S.Shenker  Nucl.Phys. {\bf B335} (1990) 635.

\noindent
7. D.Gross, A.Migdal  Phys.Rev.Lett. {\bf 64} (1990) 127.

\noindent
8. M.Douglas  Phys.Lett. {\bf B238} (1990) 635.

\noindent
9. M.Fukuma, H.Kawai, R.Nakayama  Int.J.Mod.Phys. {\bf A6} (1991) 1385.

\noindent
10. A.Belavin, A.Polyakov, A.Zamolodchikov  Nucl.Phys. {\bf B241} (1984) 333.

\noindent
11. B.Feigin, D.Fuchs  Func.Anal. \& Appl. {\bf 17} (1983) 269.

\noindent
12. Vl.Dotsenko, V.Fateev  Nucl.Phys. {\bf B240[FS12]} (1984) 312.

\noindent
13. M.Douglas, N.Seiberg, S.Shenker  Phys.Lett. {\bf B244} (1990) 381.

\noindent
14. T.Shiota  Inv.Math. {\bf 83} (1986) 333.

\noindent
15. T.Takebe  {\it From general Zakharov-Shabat equations to the
KP and the Toda lattice hierarchies}  Preprint $RIMS-779$, Kyoto (1991).

\noindent
16. E.Witten  {\it On Kontsevich model and other models of
two-dimensional gravity}  Preprint $IAASNS-HEP-91/24.$

\noindent
17. A.Marshakov, A.Mironov, A.Morozov  Phys.Lett. {\bf B274} (1992) 280.

\noindent
18. M.Kontsevich  Funk.Anal. \& Appl., {\bf 25} (1991) 50;  {\it Intersection
theory on module space of curves and the matrix Airy function}  Preprint
$MPI/91-77.$

\noindent
19. T.Miwa  Proc. of the Japan Academy, {\bf 58} (1982) 9.

\noindent
20. M.Fukuma, H.Kawai, R.Nakayama {\it Explicit solution to $p-q$ duality in
two-dimensional quantum gravity}  Preprint $UT-582, KEK-TH-289 (1991)$.

\noindent
21. R.Dijkgraaf, E.Verlinde, H.Verlinde  Nucl.Phys. {\bf B348} (1991) 435.

\noindent
22. J.Distler  Nucl.Phys. {\bf B342} (1990) 523.

\noindent
23. A.Gerasimov et.al.  Phys.Lett. {\bf B242} (1990) 345.

\noindent
24. S.Kharchev et.al., Preprint $FIAN/TD-07/92.$

\noindent
25. A.Lossev  {\it Descendants constructed from matter field and K}.{\it Saito
higher residue pairing in Landau-Ginsburg theories coupled to topological
gravity}  Preprint TPI-MINN -92/6-T.

\noindent
26. S.Kharchev et.al., to appear

\noindent
27. R.Penner  J.Diff.Geom {\bf 27} (1987) 35.

\noindent
28. J.Distler, C.Vafa  Mod.Phys.Lett. {\bf A6} (1991) 259.

\noindent
29. I.Klebanov  {\it String theory in two dimensions}  Preprint PUPT-1271,
(1991).

\noindent
30. L.Chekhov, Yu.Makeenko  {\it A hint on the external field problem for
matrix models}  Preprint NBI-HE-92-06.

\end{document}